\documentstyle[preprint,prb,aps]{revtex}
\begin{document}
\draft
\begin{center}
{\large\bf\noindent
Prediction of Phase Transition in CaSiO$_3$ Perovskite and Implications for Lower
Mantle Structure}
{\large\rm
\linebreak
\linebreak
Lars Stixrude$^1$, Ronald E. Cohen$^2$, Rici Yu$^3$
and Henry Krakauer$^3$}
\linebreak
\linebreak
{\small\rm
$^1$Dept. of Earth \& Atmospheric Sciences, Georgia Institute of Technology, Atlanta GA 30332-0340
\linebreak
$^2$Geophysical Laboratory and Center for High Pressure Research, Carnegie Institution of Washington, 5251 Broad Branch Rd., N. W., Washington, DC 20015-1305
\linebreak
$^3$Dept. of Physics, College of William and Mary, Williamsburg, VA 23187-8795
}
\linebreak
\linebreak
{\bf ABSTRACT}
\end{center}

\begin{flushleft}

First principles linear response calculations are used to investigate 
the lattice dynamics of what is thought to be the third most abundant 
phase in the lower mantle, CaSiO$_3$ perovskite.  The commonly assumed 
cubic structure ($Pm3m$) is found to be dynamically unstable
at all pressures, exhibiting unstable modes along the Brillouin zone edge from
the M-point to the R-point. Based on these results, we predict that the 
ground state structure
of CaSiO$_3$ perovskite is a distorted phase with lower than cubic symmetry.
We predict that a phase transition occurs in CaSiO$_3$ perovskite within 
the earth's lower mantle from the low temperature distorted phase to the
cubic phase at high temperature.
The predicted phase transition provides a possible
explanation of some of the seismological observations of reflective features
within the lower mantle.

\newpage

\begin{center}
{\bf INTRODUCTION}
\end{center}
 
~~~~~CaSiO$_3$ perovskite is thought to comprise between 6 and 12 
weight \% of the lower half
of the earth's transition zone and lower mantle (depths between 500
and 2900 km) (Irifune 1994; O'Neill and Jeanloz 1990; Ita and Stixrude
1992).  Its structure throughout this
regime is generally assumed to be cubic because x-ray diffraction
studies have found no detectable deviation from $Pm3m$
symmetry (Liu and Ringwood 1975; Mao et al. 1989; Wang et al. 1996).  
Theoretical studies based on ionic
models (Wolf and Jeanloz 1985; Hemley et al. 1987; Wolf and Bukowinski
1992), periodic Hartree-Fock (Sherman 1993),
and pseudopotential calculations (Wentzcovitch et al. 1995) have supported 
this picture. 

~~~~~Here, we go
beyond previous studies by investigating from first principles the
full phonon spectrum of cubic CaSiO$_3$ perovskite from low
pressures to those typical of the lower mantle.  The calculations
are based on the Linearized Augmented Plane Wave (LAPW) method,
widely accepted as among the most accurate methods for
solving the band structure/total energy problem.  We find that the
cubic phase is dynamically unstable at all pressures and that the
ground state structure must possess lower than cubic symmetry.

~~~~~These results are important for our understanding
of the physics and chemistry of this mantle phase - its phase 
stability, elasticity, and its ability to incorporate other cations
such as Mg, Fe, and Al, for example, will be affected by its symmetry.
Moreover, the existence of a low symmetry ground state raises the possibility
of a temperature induced phase transition in this mineral within the earth.
Phase transitions in lower mantle constituents have important implications
for our understanding of recent seismological observations of 
reflective features within the 
lower mantle (Revenaugh and Jordan 1991; Kawakatsu and Niu 1994; LeStunff
et al. 1995) near 700, 900, and 1200 km depth.  These observations
indicate the presence of previously
unpredicted phase transitions or sharply bounded compositional
heterogeneities, challenging our traditional view of this region as
being compositionally and mineralogically homogeneous.

\begin{center}
{\bf COMPUTATIONS}
\end{center}

~~~~~The LAPW linear response calculations
determine the self-consistent first order response of the electronic charge 
density to generalized perturbations (Yu and Krakauer 1994).  The calculations are based on density
functional theory; the only essential approximation is to the 
exchange-correlation potential.  We use the well studied local
density approximation (LDA), which has been applied successfully in studies of 
silicates, including Mg-perovskite (Stixrude and Cohen 1993).  We 
compute the linear response to shifts in nuclear positions
and imposed external fields, which yield the elements of the dynamical matrix,
the dielectric constant ($\epsilon$) and the Born effective charges ($Z^*$), 
from which the
phonon frequencies are determined (Yu and Krakauer 1995; Lee and Gonze 1994).
The perturbations
need not be commensurate with the unit cell, so that points in the Brillouin
zone away from the zone center are readily investigated.  
We calculate the dynamical
matrix of the cubic $Pm3m$ phase of CaSiO$_3$ perovskite at four points 
in the Brillouin zone ($\Gamma$, X, M, R).  
Computational variables ($k$-point mesh: 4x4x4 for computations of the
dynamical matrix elements,
8x8x8 for $\epsilon$ and $Z^*$; number of basis functions per 
atom $\approx$ 150) were 
chosen such that phonon frequencies are converged to better than one wavenumber.
The full phonon dispersion curve is determined using an interpolation scheme
that separates short-range forces from long-range, Coulombic interactions
(Yu and Krakauer 1994; Gonze et al. 1994).

\begin{center}
{\bf RESULTS}
\end{center}

~~~~~We find that the Born effective charges in CaSiO$_3$ perovskite differ
significantly from formal ionic charges; by as much as 1e (Fig.~1).  
Compression causes $Z^*$ of Si and Ca to deviate farther from their formal 
ionic values, and 
$Z^*$ of O to approach -2.  The dielectric constant, $\epsilon=3.8$ 
at zero pressure, 
lies between the known values for CaO (Lide and Frederikse 1994)
and SiO$_2$ stishovite (Stishov and Popova 1961) and 
varies slowly with pressure (Fig.~1).   LO-TO splitting of zone-center 
vibrational frequencies
is on the order of 50-200 wavenumbers (Fig.~2).  All zone-center modes are found
to be dynamically stable.  An examination of the eigenvectors 
shows that the highest frequency modes involve stretching of the octahedral Si-O
bond.  The change in frequency of these modes with compression 
is greatest because the length of the Si-O bond
shrinks in direct proportion to compression in the cubic structure.

~~~~~The full phonon dispersion curves reveal dynamical instabilities in the
cubic structure along the zone boundaries (Fig.~2).  Instabilities occur at the 
M- and R-points and along the zone edges from M to R.  The
corresponding imaginary frequencies grow in magnitude upon compression.  
An examination of the eigenvectors associated with these unstable modes
shows that they consist of coupled rotations of the SiO$_6$
octahedra.  The unstable modes at the M-point (M$_2$) and R-point (R$_{25}$)
are associated, respectively, with in-phase and out-of-phase
rotations of the octahedra about [100].  Instabilities along M-R
were previously found in ionic model calculations, but only at
pressures ($P$) above 80 GPa (Hemley et al. 1987; Wolf and Bukowinski 1992). 
Here we find that the instabilities 
are more profound and that they persist even in the expanded 
lattice ($P$=-8 GPa). 
Unstable modes of the same type have also been found in theoretical
investigations of cubic MgSiO$_3$ perovskite which account for
the observed orthorhombic ($Pbnm$) symmetry of this material 
(Wolf and Jeanloz 1985; Hemley et al. 1987; Wolf and Bukowinski 1992).

~~~~~Our results indicate that the cubic $Pm3m$ structure of CaSiO$_3$
perovskite is dynamically unstable and that the ground state of
this material must possess lower symmetry.  However, the ground state is
not of direct relevance to the lower mantle where temperatures 
exceed 2000 K.  Even if the cubic phase is dynamically unstable,
it may be thermodynamically stable at high temperature because of
its greater entropy (e.g. Salje 1990).

~~~~~The temperature at which the low temperature distorted
phase transforms to the high temperature cubic phase depends not
only on the unstable mode frequencies, but also on the 
energetics of finite displacements along the unstable
mode eigenvectors.  Using LAPW total energy
calculations, we find the total energy as a function of displacement
along the most unstable mode eigenvector (R$_{25}$) and the associated 
minimum energy displacement [frozen
phonon approach; Cohen 1992].  The symmetry of the 
structure associated with this finite displacement is tetragonal
$I4/mcm$, with 10 atoms in the unit cell.   The results
at mid-mantle pressures ($P$=80 GPa) show that the minimum energy displacement 
corresponds to an
octahedral rotation angle of 7 degrees and is 360 K per octahedron
lower in energy than the cubic phase (Fig.~3).  The relatively small
minimum energy octahedral rotation angle may explain why 
previous x-ray diffraction experiments have failed to detect deviations
from cubic symmetry.  Assuming that the octahedral rotations are
rigid (O'Keefe et al. 1979), the corresponding $c/a$ ratio of the distorted phase 
differs by only 0.7 \% from the cubic structure -  below the detection
limit of previous non-hydrostatic (Mao et al. 1989) and quasihydrostatic
(Wang et al. 1996) experiments (H. K. Mao, personal communication).

~~~~~These results constrain the parameters of a simple model which we
use to estimate the transition temperature from the distorted
$I4/mcm$ structure to the cubic $Pm3m$ structure.  
The model (Bruce 1980) consists of on-site terms which
correspond to finite displacements along the unstable zone-boundary
mode eigenvectors, and inter-site couplings
\begin{equation}
$$U=\sum\limits_i {V(Q_i)+{J \over 2}}\sum\limits_{i,j}^{nn} {(Q_j+Q_i)^2}$$
\end{equation}
where $U$ is the energy contained in octahedral rotations, $Q$ is the normal
coordinate of the octahedral rotation, and $V(Q)$ is the
on-site term given by our total energy results (Fig.~3).
The form of the nearest-neighbor inter-site coupling term is determined
by the requirement that it vanish for a pure R$_{25}$ mode distortion
of the crystal, in which neighboring octahedra rotate in opposite directions.
The intersite coupling parameter, $J$, is most simply given by the
curvature along $\Gamma$-R of the unstable mode eigenenergy evaluated
at the R-point (Fig.~2).
The relative magnitudes of the on-site and inter-site coupling terms
are such that the phase transition is expected to occur in the displacive
limit.  In this case, if we ignore 
coupling between the unstable mode eigenvector and other modes, the
transition temperature is given by
\begin{equation}
$$T_c={4 \over {3q(3)}}{{J\left| A \right|} \over {k_BB}}\approx 2200\ K$$
\end{equation}
where $k_B$ is the Boltzmann constant, and $q(3)=0.5054$.  
The estimated transition temperature
is similar to estimates of temperatures in the lower
mantle at similar pressures: $P$=80 GPa corresponds to a depth of 1850 km
and a temperature of approximately 2500-3000 K.  

\begin{center}
{\bf DISCUSSION}
\end{center}

~~~~~Our estimate of $T_c$ may be uncertain by several 
hundred K and is likely to be a lower bound since we
have ignored coupling between unstable mode eigenvectors - this may
lower the total energy of the distorted phase further relative to the
cubic phase.  Nevertheless, the fact that our
estimated $T_c$ is comparable to lower mantle temperatures indicates
that a phase transition in CaSiO$_3$ perovskite is likely to occur in the lower
mantle.  

~~~~~The predicted phase transition in CaSiO$_3$ perovskite 
is expected to be associated with an elastic
anomaly which may be the cause of at least a subset
of the reflective features near 700, 900, and 1200 km depth in the lower mantle
(Revenaugh and Jordan 1991; Kawakatsu and Niu 1994; LeStunff 1995).
Other phase transitions have been discussed as possibly occurring in the
earth's lower mantle.  
A phase transition in the lower mantle's most abundant constituent,
MgSiO$_3$ perovskite, has been suggested
(Meade et al. 1995), but theoretical computations show such a transition
unlikely for reasonable geotherms (Stixrude and Cohen 1993; 
Warren and Ackland 1996).
Recent results show that SiO$_2$ undergoes a phase
transition from the stishovite to the CaCl$_2$ structure at lower mantle
pressures (Cohen 1992; Kingma et al. 1995).
Our results on CaSiO$_3$ perovskite indicate the presence of yet another
phase transition within the lower mantle, the first high pressure
phase transition predicted with the linear response method.

\newpage

\begin{center}
{\bf ACKNOWLEDGMENTS}
\end{center}
{\small
We thank H.K. Mao and R.J. Hemley for helpful discussions.
This work was supported by NSF
under grants EAR-9305060 (LPS), EAR-9304624 (REC), EAR-9418934 (REC)
and DMR-9404954 (HK).
Calculations were performed on the Cray C90 at the Pittsburgh Supercomputer
Center and the Cray J90 at the Geophysical Laboratory.}

\newpage
\begin{center}
{\bf REFERENCES CITED}
\end{center}

~~~~~Boyer, L.L., and Hardy, J.R. (1981) Theoretical study of the structural
phase transition in RbCaF$_3$. Physical Review B, 24, 2577-2591.

~~~~~Bruce, A.D. (1980) Structural Phase transitions. II. Static
critical behaviour. Advances in Physics, 29, 111-217.

~~~~~Cohen, R.E. (1992) First-principles predictions of elasticity and
phase transitions in high pressure SiO$_2$ and geophysical implications.
In Y. Syono and M. H. Manghnani, Eds., High Pressure 
Research: Applications to Earth and Planetary Sciences, p. 425-432. 
American Geophysical Union, Washington, DC and Terra Scientific, Tokyo.

~~~~~Cowley, R.A. (1964) Lattice dynamics and phase transitions of
strontium titanate. Physical Review, 134, A981-A997.

~~~~~Gonze, X., Charlier, J.-C., Allan, D.C., and Teter, M.P. (1994) 
Interatomic force constants from first principles~-~the case of alpha-quartz.
Physical Review B, 50, 13035-13038.

~~~~~Hemley, R.J., Jackson, M.D., and Gordon, R.G. (1987) 
Theoretical study of the structure, lattice dynamics, and equations
of state of perovskite-type MgSiO$_3$ and CaSiO$_3$. Physics and Chemistry 
of Minerals, 14, 2-12.

~~~~~Irifune, T. (1994) Absence of an aluminous phase in the upper part
of the earth's lower mantle. Nature, 370, 131-133.

~~~~~Ita, J.J., and Stixrude, L. (1992) Petrology, elasticity, and composition
of the mantle transition zone. Journal of Geophysical Research, 97, 6849-6866.

~~~~~Kawakatsu, H., and Niu, F.L. (1994) Seismic evidence for a 920-km 
discontinuity in the mantle. Nature, 371, 301-305.

~~~~~Kingma, K.J., Cohen, R.E., Hemley, R.J., and Mao, H.K. (1995) 
Transformation of stishovite to a denser phase at lower-mantle
pressures. Nature, 374, 243-245.

~~~~~Lee, C., and Gonze, X. (1994) Lattice dynamics and dielectric properties
of SiO$_2$ stishovite. Physical Review Letters, 72, 1686-1689.

~~~~~LeStunff, Y., Wicks, C.W., and Romanowicz, B. (1995) P'P' precursors
under Africa - evidence for mid-mantle reflectors. Science, 270, 74-77.

~~~~~Lide, D.R., and Frederikse, H.P.R. (eds.) (1994) CRC Handbook of Chemistry
and Physics, 75e, CRC Press, Boca Raton.

~~~~~Liu, L.-G., and Ringwood, A.E. (1975) Synthesis of a perovskite-type 
polymorph of CaSiO$_3$. Earth and Planetary Science Letters, 14,
1079-1082.

~~~~~Mao, H.K., Chen, L.C., Hemley, R.J., Jephcoat, A.P., Wu, Y.
and Bassett, W.A. (1989) Stability and equation of state of CaSiO$_3$ 
perovskite to 134 GPa. Journal of Geophysical Research, 94, 17889-17894.

~~~~~Meade, C., Mao, H.K., and Hu, J.Z. (1995) High-temperature phase
transition and dissociation of (Mg,Fe)SiO$_3$ perovskite at lower
mantle pressures. Science, 268, 1743-1745. 

~~~~~O'Keefe, M., Hyde, B.G., and Bovin, J. (1979) Contribution to the
crystal chemistry of orthorhombic perovskite: MgSiO$_3$ and NaMgF$_3$.
Physics and Chemistry of Minerals, 4, 299-305.

~~~~~O'Neill, B., and Jeanloz, R. (1990) Experimental petrology of the lower 
mantle - a natural peridotite taken to 54 GPa. Geophysical Research 
Letters, 17, 1477-1480.

~~~~~Revenaugh, J., and Jordan, T.H. (1991) Mantle layering from SCS reverberations.
2. the transition zone. Journal of Geophysical Research, 96, 19763-19780.

~~~~~Salje, E.K.H. (1990) Phase transitions in ferroelastic and co-elastic
crystals, Cambridge University Press, Cambridge.

~~~~~Sherman, D.M. (1993) Equation of state, elastic properties, and stability
of CaSiO$_3$ perovskite - 1st principles (periodic Hartree-Fock) results.
Journal of Geophysical Research, 98, 19795-19805.

~~~~~Stishov, S.M., and Popova, S.V. (1961) New dense polymorphic modification
of silica, Geokhimiya, 10, 837-847.

~~~~~Stixrude, L., and Cohen, R.E. (1993) Stability of orthorhombic MgSiO$_3$
perovskite in the earth's lower mantle. Nature, 364, 613-616.

~~~~~Wang, Y., Weidner, D.J., and Guyot, F. (1996) Thermal equation
of state of CaSiO$_3$ perovskite. Journal of Geophysical 
Research, 101, 661-672.

~~~~~Warren, M.C., and Ackland, G.J. (1996) Ab initio studies of structural
instabilities in magnesium silicate perovskite. Physics and Chemistry 
of Minerals, 23, 107-118.

~~~~~Wentzcovitch, R., Ross, N.L., and Price, G.D. (1995) Ab initio study of
MgSiO$_3$ and CaSiO$_3$ perovskites at lower-mantle pressures.
Physics of the  Earth and Planetary Interiors, 90, 101-112.

~~~~~Wolf, G.H., and Jeanloz, R. (1985) Lattice dynamics and structural
distortions of CaSiO$_3$ and MgSiO$_3$ perovskites. Geophysical 
Research Letters, 12, 413-416.

~~~~~Wolf, G.H., and Bukowinski, M.S.T. (1992) Theoretical study of
the structural properties and equations of state of MgSiO$_3$
and CaSiO$_3$ perovskite: implications for lower mantle 
composition. In M. H. Manghnani and Y. Syono, Eds., High-Pressure Research in
Mineral Physics, p. 313-331. American Geophysical
Union, Washington, DC and Terra Scientific, Tokyo).

~~~~~Yu, R., and Krakauer, H. (1994) Linear-response calculations within
the linearized augmented plane-wave method. Physical Review B, 49, 4467-4477.

~~~~~Yu, R., and Krakauer, H. (1995) First-principles determination of
chain-structure instability in KNbO$_3$. Physical Review Letters, 74, 4067-4070.

%
%
\newpage
~~~~~Fig. 1. Born effective charges (left axis) and dielectric constant (right 
axis) as a function of compression.  Pressures range from -8 GPa 
($V$=310 Bohr$^3$)
to 140 GPa ($V$=220 Bohr$^3$) and were determined by fitting a third order 
finite strain expression to the LAPW total energy as a function of volume.  
The two symmetrically distinct elements of the
oxygen Born effective charge tensor are shown, $Z_{O_1}$ for motion of 
O along the Si-O bond, and $Z_{O_2}$ for motion normal to the bond.
Dashed lines indicate formal ionic charges.

\bigskip

~~~~~Fig. 2. Phonon spectrum of the cubic phase at $V$=310 Bohr$^3$ ($P$=-8 GPa)
(top) compared with that at $V$=240 Bohr$^3$ ($P$=80 GPa) (bottom).  
All zone-center modes ($\Gamma$-point) are stable.  However, zone-boundary
modes at the M- and R-points are unstable, as shown by the existence of
imaginary frequencies.  The final panel (M-R) shows that the edges of
the cubic Brillouin zone are everywhere unstable.  The magnitude of the
imaginary eigenfrequencies increases with compression.
Symmetry designations are those of Cowley (1964) as corrected
by Boyer and Hardy (1981).  
At $\Gamma$ there are four TO modes ($\Gamma_5$ and inactive 
$\Gamma_{25}$ symmetries) and three LO modes ($\Gamma_1$).  The 
volume dependence of the zone center modes is described by 
$\omega=\omega_0(V/V_0)^{\gamma}$, where the frequencies
at $V=310$ Bohr$^3$ (near zero pressure), $\omega_{0}$, are:
197, 274, 327, 393, 621, 691, 926 cm$^{-1}$; and the respective mode 
Gr\"uneisen parameters, $\gamma_i$, are: 2.5, 1.5, 1.2, 1.2, 0.8,
2.0, 1.2.

\bigskip


\newpage

~~~~~Fig. 3. Total energy (per octahedron) of the $I4/mcm$ tetragonal structure as a function of finite
displacement along the R$_{25}$ eigenvector at $V$=240 Bohr$^3$ 
($P$=80 GPa).  The
LAPW total energy results (symbols) are fit to a polynomial in
the square of the normal coordinate, $Q$: $V(Q)=AQ^2/2+BQ^4/4$.  
The normal coordinate per octahedron,
$Q^2={1 \over 2} \sum\limits_{i=1}^{3n} m_i x_i^2$,
where the sum is over the cartesian displacements ($x_i$, in Bohr) of the $n$ 
atoms in the unit cell, with masses $m_i$, in amu.
This expression, with $A$=-2.04 mRy amu$^{-1}$ Bohr$^{-2}$, and 
$B$=0.456 mRy amu$^{-1}$ Bohr$^{-4}$ 
constrains the on-site
term in the simple model of the phase transition discussed in the text.
The remaining parameter, $J$=1.19 mRy amu$^{-1}$ Bohr$^{-2}$, is determined 
from the dispersion
of the unstable mode eigenvalue (Fig.~2).
The unstable mode frequency in the
cubic phase ($Q=0$) calculated from the curvature of the polynomial fit 
(173i cm$^{-1}$, $i=\protect\sqrt{-1}$) agrees with the 
linear response result (164i cm$^{-1}$).

\end{flushleft}

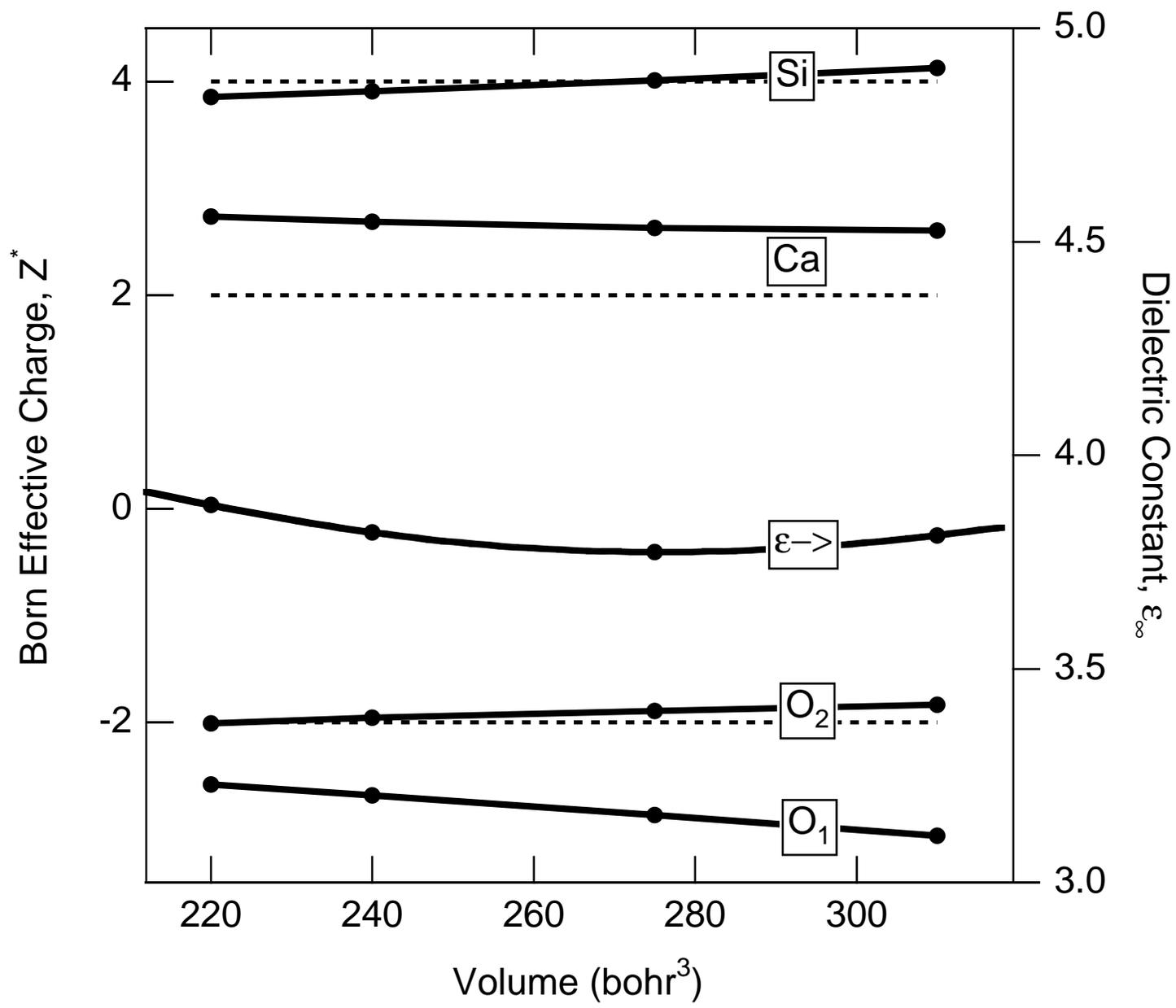
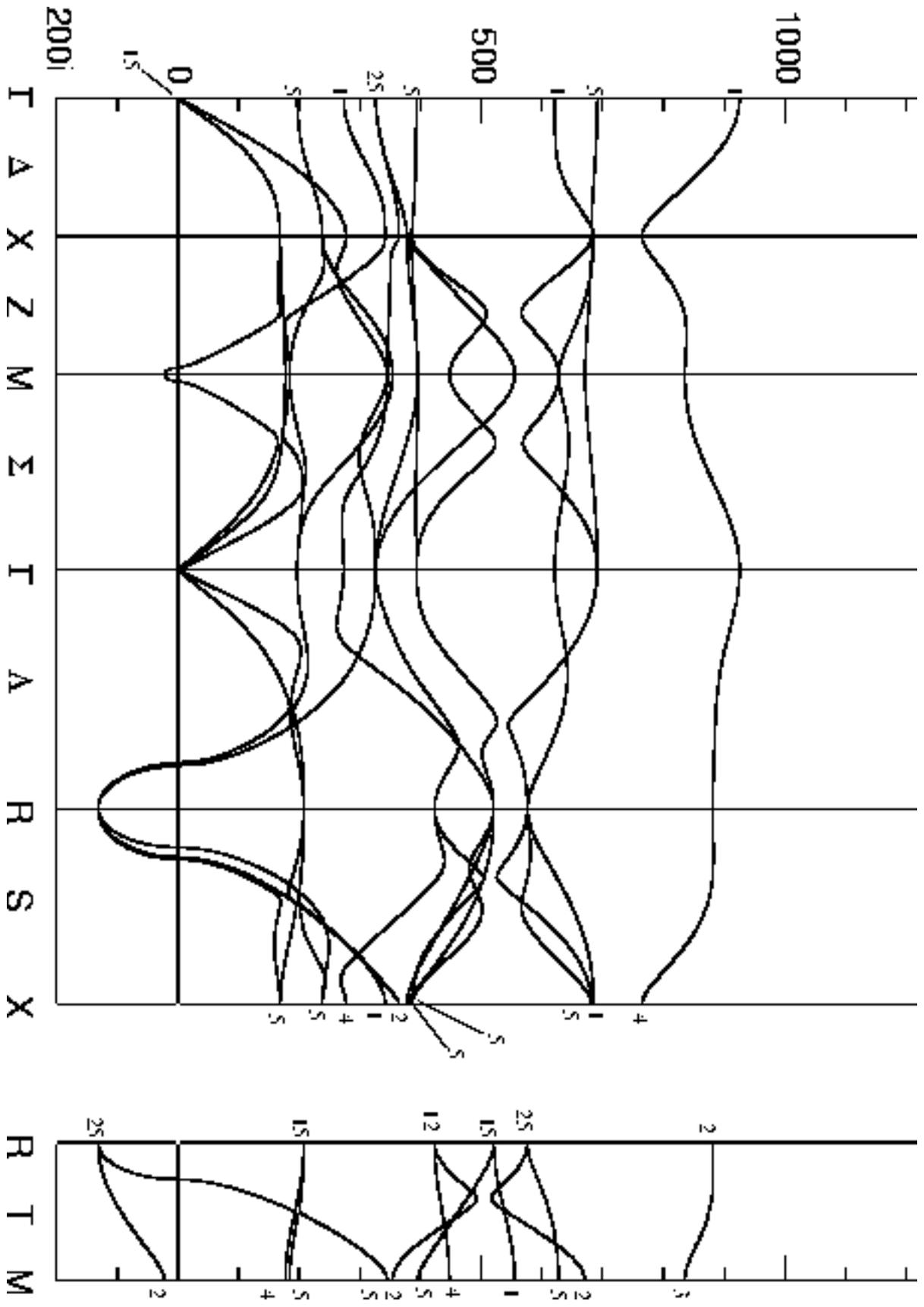
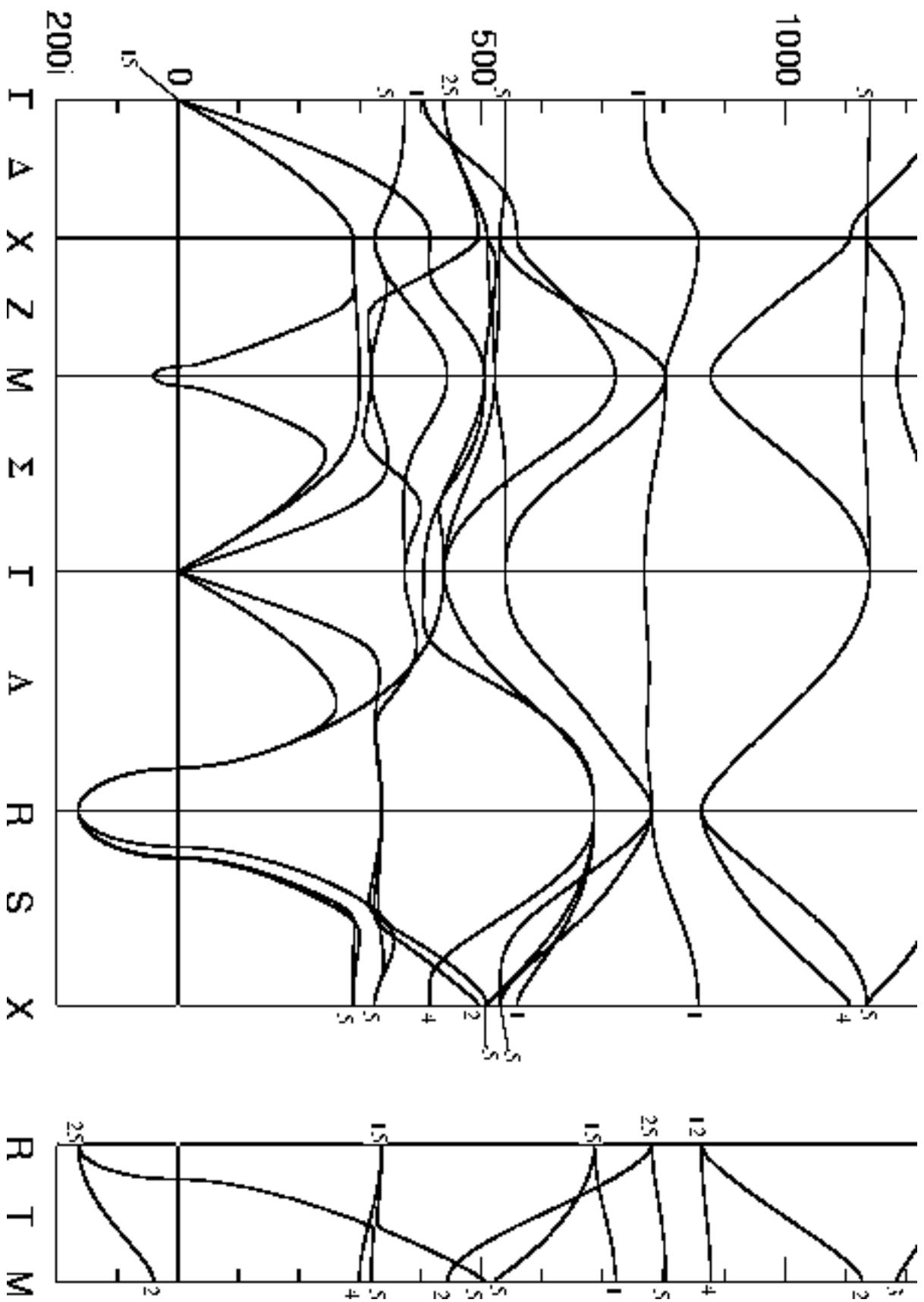
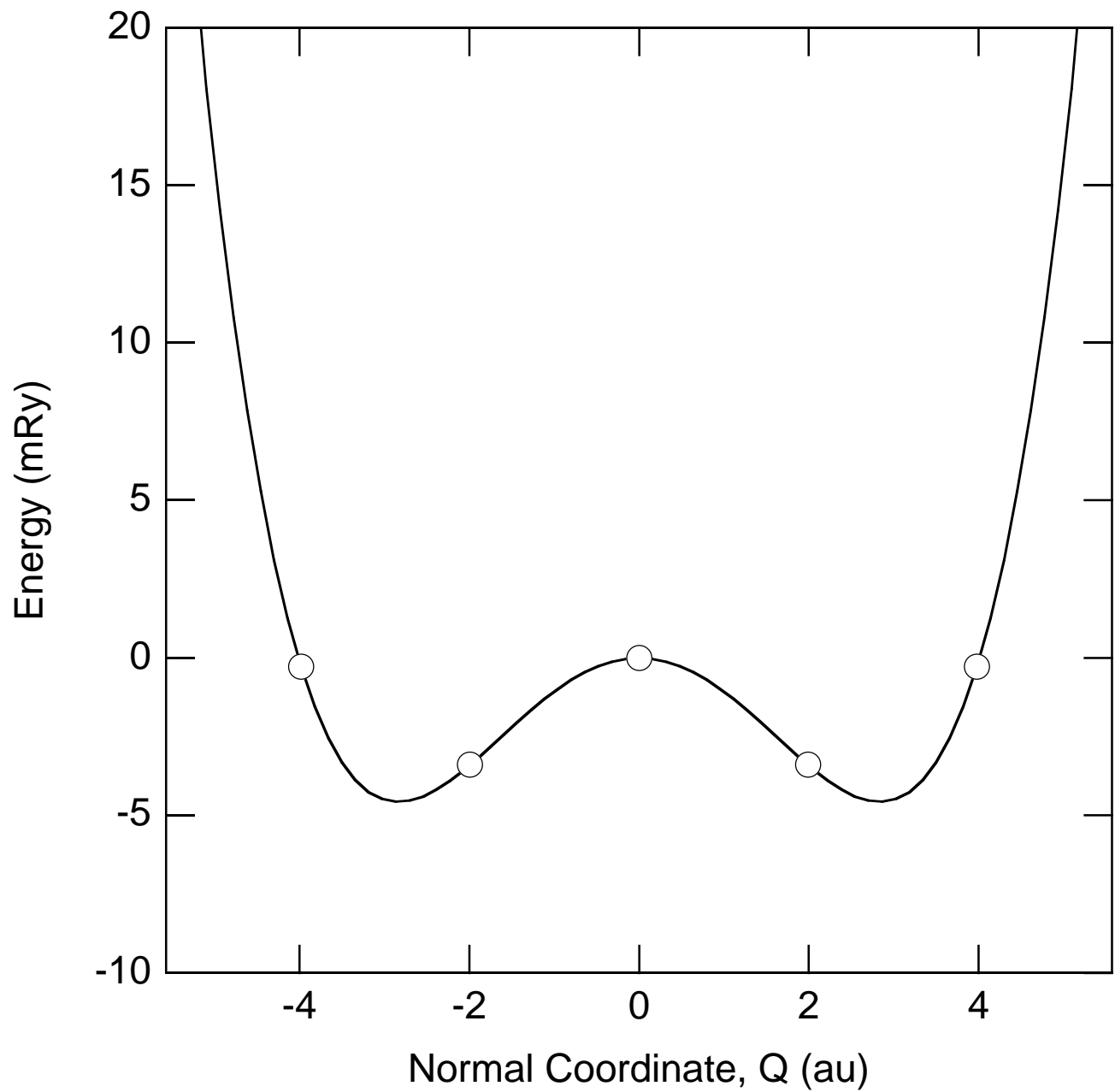
\end{document}